\documentclass[a4paper,fleqn]{cas-sc}

\usepackage[authoryear,longnamesfirst]{natbib}
\usepackage{subfigure}
\usepackage{algorithm,algpseudocode,multicol}
\usepackage{float}
\usepackage{soul} 

\usepackage{diagbox}
\def\tsc#1{\csdef{#1}{\textsc{\lowercase{#1}}\xspace}}
\tsc{WGM}
\tsc{QE}
\tsc{EP}
\tsc{PMS}
\tsc{BEC}
\tsc{DE}
\begin{document}
	%\begin{sloppypar}
\let\WriteBookmarks\relax
\def\floatpagepagefraction{1}
\def\textpagefraction{.001}

\newtheorem{thm}{Theorem}
\newtheorem{lem}{Lemma}
\newtheorem{cor}{Corollary}
\newtheorem{rem}{Remark}
\newtheorem{assum}{Assumption}
\newtheorem{prob}{Problem}
\newtheorem{defn}{Definition}
\newproof{pf}{Proof}
\newproof{pot}{Proof of Theorem \ref{thm2}}

% Short title
\shorttitle{Trajectory Optimization for UAV-Based Medical Delivery}

% Short author
\shortauthors{Chen et~al.}

% Main title of the paper
\title [mode = title]{Trajectory Optimization for UAV-Based Medical Delivery with Temporal Logic Constraints and Convex Feasible Set Collision Avoidance}
%{How to Divide: A Set Partitioning Strategy Balancing the Trade-off Between Intra-Subset Correlation and Inter-Subset Gain Mutual Influence in Distributed Attack Detection Scheduling Task}                      
% Title footnote mark
% eg: \tnotemark[1]
%\tnotemark[1,2]
%
%% Title footnote 1.
%% eg: \tnotetext[1]{Title footnote text}
%% \tnotetext[<tnote number>]{<tnote text>} 
%\tnotetext[1]{This document is the results of the research
%   project funded by the National Science Foundation.}
%
%\tnotetext[2]{The second title footnote which is a longer text matter
%   to fill through the whole text width and overflow into
%   another line in the footnotes area of the first page.}

% First author
%
% Options: Use if required
% eg: \author[1,3]{Author Name}[type=editor,
%       style=chinese,
%       auid=000,
%       bioid=1,
%       prefix=Sir,
%       orcid=0000-0000-0000-0000,
%       facebook=<facebook id>,
%       twitter=<twitter id>,
%       linkedin=<linkedin id>,
%       gplus=<gplus id>]
\author[1]{Kaiyuan Chen}[style=chinese]
%\ead{kaiyuanchen@mail.tsinghua.edu.cn}
\credit{Methodology, Software, writing-original draft, Writing - Review \& Editing}
% Corresponding author indication

\affiliation[1]{organization={State Key Laboratory of Multimodal Artificial Intelligence Systems},
	addressline={Institute of Automation, Chinese Academy of Sciences}, 
	city={Beijing},
	% citysep={}, % Uncomment if no comma needed between city and postcode
	postcode={100190}, 
	% state={},
	country={China}}
    
\affiliation[2]{organization={Vanke School of Public Health},
	addressline={Tsinghua University}, 
	city={Beijing},
	% citysep={}, % Uncomment if no comma needed between city and postcode
	postcode={100084}, 
	% state={},
	country={China}}
    
\author[2]{Yuhan Suo}[style=chinese]
%[type=editor,
%                        auid=000,bioid=1,
%                        prefix=Sir,
%                        role=Researcher,
%                        orcid=0000-0001-7511-2910]

% Footnote of the first author
%\fnmark[1]

% Email id of the first author
%\ead{yuhan.suo@bit.edu.cn}
\credit{Conceptualization, Methodology, Software, writing-original draft}
%\credit{Data curation, Writing - Original draft preparation}
% URL of the first author
%\ead[url]{www.cvr.cc, cvr@sayahna.org}

%  Credit authorship
%\credit{Conceptualization of this study, Methodology, Software}

% Address/affiliation
\affiliation[3]{organization={School of Automation},
    addressline={Beijing Institute of Technology}, 
    city={Beijing},
    % citysep={}, % Uncomment if no comma needed between city and postcode
    postcode={100081}, 
    % state={},
    country={China}}

% Third author
\author[2]{Shaowei Cui}[style=chinese]
%\fnmark[2]
%\ead{xia_yuanqing@bit.edu.cn}
%\ead[URL]{www.sayahna.org}
\credit{Writing - Review \& Editing}

% Third author
\author[2]{Yuanqing Xia}[style=chinese]
%\fnmark[2]
%\ead{xia_yuanqing@bit.edu.cn}
%\ead[URL]{www.sayahna.org}
\credit{Review, supervision}

\author[3]{Wannian Liang}[style=chinese]
%\fnmark[2]
\ead{liangwn@
	tsinghua.edu.cn}
%\ead[URL]{www.sayahna.org}
\credit{Review, supervision}
\cormark[1]

% Third author
\author[1]{Shuo Wang}[style=chinese]
%\fnmark[2]
\ead{shuo.wang@ia.ac.cn}
%\ead[URL]{www.sayahna.org}
\credit{Writing - Review \& Editing}
\cormark[1]
% Address/affiliation
%\affiliation[2]{organization={Sayahna Foundation},
%    % addressline={}, 
%    city={Jagathy},
%    % citysep={}, % Uncomment if no comma needed between city and postcode
%    postcode={695014}, 
%    state={Trivandrum},
%    country={India}}

%\author[4]{Jiping Xu}[style=chinese]
%%\fnmark[2]
%\ead{xujp@th.btbu.edu.cn}
%%\ead[URL]{www.sayahna.org}
%\credit{Review}
%% Address/affiliation
%%\affiliation[2]{organization={Sayahna Foundation},
%	%    % addressline={}, 
%	%    city={Jagathy},
%	%    % citysep={}, % Uncomment if no comma needed between city and postcode
%	%    postcode={695014}, 
%	%    state={Trivandrum},
%	%    country={India}}
%\affiliation[4]{organization={School of Computer and Artificial Intelligence},
%	addressline={Beijing
%		Technology and Business University}, 
%	city={Beijing},
%	% citysep={}, % Uncomment if no comma needed between city and postcode
%	postcode={100048}, 
%	% state={},
%	country={China}}

% Fourth author
% \author%
% [1]
% {Yuanqing Xia}[style=chinese]
% %\cormark[2]
% %\fnmark[1,3]
% \ead{xia_yuanqing@bit.edu.cn} 
% \credit{Review}
%\ead[URL]{www.stmdocs.in}

%\affiliation[3]{organization={STM Document Engineering Pvt Ltd.},
%    addressline={Mepukada}, 
%    city={Malayinkil},
%    % citysep={}, % Uncomment if no comma needed between city and postcode
%    postcode={695571}, 
%    state={Trivandrum},
%    country={India}}

% Corresponding author text
\cortext[cor1]{Corresponding author}
%\cortext[cor2]{Principal corresponding author}

%% Footnote text
%\fntext[fn1]{This is the first author footnote. but is common to third
%  author as well.}
%\fntext[fn2]{Another author footnote, this is a very long footnote and
%  it should be a really long footnote. But this footnote is not yet
%  sufficiently long enough to make two lines of footnote text.}
%
%% For a title note without a number/mark
%\nonumnote{This note has no numbers. In this work we demonstrate $a_b$
%  the formation Y\_1 of a new type of polariton on the interface
%  between a cuprous oxide slab and a polystyrene micro-sphere placed
%  on the slab.
%  }

% Here goes the abstract
\begin{abstract}
This paper addresses the problem of trajectory optimization for unmanned aerial vehicles (UAVs) performing time-sensitive medical deliveries in urban environments. Specifically, we consider a single UAV with 3-degree-of-freedom dynamics tasked with delivering blood packages to multiple hospitals, each with a predefined time window and priority. Mission objectives are encoded using Signal Temporal Logic (STL), enabling the formal specification of spatial-temporal constraints. To ensure safety, city buildings are modeled as 3D convex obstacles, and obstacle avoidance is handled through a Convex Feasible Set (CFS) method. The entire planning problem—combining UAV dynamics, STL satisfaction, and collision avoidance—is formulated as a convex optimization problem that ensures tractability and can be solved efficiently using standard convex programming techniques. Simulation results demonstrate that the proposed method generates dynamically feasible, collision-free trajectories that satisfy temporal mission goals, providing a scalable and reliable approach for autonomous UAV-based medical logistics.
	%In view of the problem that dynamic propagation attacks seriously threaten the security of the actual network environment, this paper explores the feasibility of actively approaching and repairing the attacked sensors to achieve targeted handling of dynamic attacks. Based on mobile sink nodes with both mobile and sniffing information capabilities, the  active response mechanism for dynamic propagation attack based on detection domain is proposed. This mechanism allows mobile sink nodes to trigger corresponding actions based on whether the attacker is in a monitorable state. In the case where mobile sink nodes lack complete confidence information about the unmonitorable attacker, the detection algorithm for unmonitorable attacks  based on partial confidence space is proposed. Theoretical analysis shows that when specific conditions are satisfied, the proposed algorithm can effectively track and detect the unmonitorable attack. Simulation results in different scenarios verify the effectiveness of the proposed algorithm.

%In view of the more threatening dynamic propagation attacks in the network environment, this paper explores the feasibility of actively approaching and repairing the attacked sensors through mobile sink nodes to achieve targeted handling of dynamic attacks.
%针对网络环境中更具威胁的动态传播攻击，本文探讨了通过移动汇聚节点主动靠近并修复受攻击传感器的可行性，以实现对动态攻击的针对性处置。
\end{abstract}

% Use if graphical abstract is present
% \begin{graphicalabstract}
% \includegraphics{figs/grabs.pdf}
% \end{graphicalabstract}

 %Research highlights
%\begin{highlights}
%	%面对更加隐蔽的自传播攻击，创新性地用移动汇聚节点动态跟踪攻击者。
%	%对于不可监控攻击，提出一种能够仅依据部分置信信息的移动策略设计算法。
%	%给出了移动汇聚节点能够在有限次移动后跟踪到任意攻击者的一系列确定条件。
%
%\item 	The detection problem of dynamic propagation attacks in the network is transformed into the problem of tracking dynamic attacks using mobile sink nodes.
%
%\item  The mobile strategy design algorithm based on partial confident information is proposed to track the unmonitorable attack.
%
%\item A series of deterministic conditions are given for the mobile sink node to track any attacker after a limited number of movements.
%\end{highlights}

% Keywords
% Each keyword is seperated by \sep

\maketitle
\setlength{\parindent}{2em} 
\section{Introduction}

Unmanned aerial vehicles are emerging as a transformative solution for rapid and reliable medical logistics, especially in emergency scenarios where timely delivery of critical supplies—such as blood, vaccines, or organs—can directly impact patient outcomes. Traditional ground transportation is often limited by traffic congestion, geographic barriers, or disaster-related disruptions. In contrast, UAVs offer the ability to bypass such constraints, enabling swift point-to-point delivery in both urban and rural settings. The potential for autonomous aerial systems to support healthcare infrastructure is particularly promising in time-critical missions such as blood transfusion logistics, where every minute counts \citep{Bauer21development}. By optimizing their navigation and delivery efficiency, UAVs can serve as a crucial component in future smart healthcare systems.

Despite their promise, deploying UAVs in dense urban environments presents significant challenges. First, UAV trajectories must respect the underlying vehicle dynamics, including acceleration, velocity, and turning limitations, to ensure feasibility and controllability in real-time flight. Second, urban settings are filled with static obstacles such as buildings, posing serious safety risks if not properly modeled and avoided \citep{Kuang24convergence}. Third, the mission objectives are often not limited to reaching a single location but require the UAV to visit multiple destinations—each with individual priorities and strict time windows \citep{Foehn21time,Romero22model}. Finally, trajectory planning must reconcile these spatial and temporal requirements with safety constraints, all while remaining computationally tractable for online or near-real-time deployment. Addressing these challenges calls for a planning framework that integrates high-level task specification with low-level dynamic and environmental constraints \citep{Szmuk19real}.

Early approaches to UAV trajectory planning have primarily focused on geometric or graph-based methods, such as A* search, rapidly-exploring random tree, and their numerous variants \citep{Funk23orientation,Aggarwal20path}. These methods are effective in generating collision-free paths in static environments but often neglect dynamic feasibility, control limits, or time-critical objectives. Moreover, they typically rely on post-processing to smooth or adapt the path, which may violate safety or dynamic constraints when applied directly to real vehicles \citep{Funk23orientation}.

To address the limitations of purely geometric planning, optimal control and trajectory optimization methods have been introduced. These approaches formulate trajectory generation as a continuous optimization problem, incorporating system dynamics and control constraints directly into the planning process \citep{Chen23simultaneous,Morrell18comparison,Betts10}. While powerful, they often require carefully tuned cost functions and may struggle to represent complex high-level tasks such as sequential visits or deadline-based deliveries \citep{Morrell18comparison}.

More recently, temporal logic frameworks, especially signal temporal logic, have been adopted to encode complex task specifications in a formal and interpretable manner \citep{Gilpin21smooth,Belta19formal}. STL enables the expression of spatial-temporal goals such as “reach a region within a certain time window,” making it suitable for high-level mission planning. However, directly incorporating STL constraints into optimization often leads to non-convex or mixed-integer formulations \citep{Buyukkocak21planning}, which can be computationally expensive and difficult to scale, especially in real-time applications.

To mitigate these challenges, recent research has explored convex relaxations and smooth approximations of STL semantics \citep{Buyukkocak25resilient,Gilpin21smooth}. By reformulating STL conditions using auxiliary variables and convex functions, it becomes possible to embed temporal logic specifications into a continuous optimization framework without sacrificing tractability. Nevertheless, integrating such methods with realistic dynamics and obstacle avoidance in three-dimensional urban settings remains an open challenge, particularly when multiple delivery targets and time-sensitive constraints are involved.

While prior works have advanced individual aspects of UAV path planning—such as geometric routing, trajectory optimization, or temporal logic-based task encoding—few have addressed the full integration of high-level temporal specifications, low-level vehicle dynamics, and obstacle-rich environments within a unified, tractable framework. Most approaches either simplify the environment to two dimensions \citep{Buyukkocak25resilient,Gundana21event}, assume unconstrained dynamics \citep{Buyukkocak21planning,Chen25fast}, or rely on computationally expensive non-convex formulations that are unsuitable for real-time applications \citep{Zhang22uav}. In contrast, our work bridges the gap between formal mission specification and practical trajectory generation by combining temporal logic, convex obstacle avoidance, and physically consistent UAV motion planning in three-dimensional urban space.

This paper presents a trajectory optimization framework tailored for UAV-based blood delivery missions under temporal and environmental constraints. The main contributions of this work are as follows:
\begin{enumerate}
	\item We propose an STL-based formulation of multi-goal medical delivery tasks, allowing each hospital visit to be precisely specified with spatial and temporal bounds.
	\item Obstacle avoidance is addressed using the convex feasible set method, which iteratively constructs linearized safe regions around a reference trajectory, enabling scalable and collision-free path planning in complex 3D environments.
	\item The entire planning problem—including STL satisfaction, obstacle avoidance, and UAV dynamics—is formulated as a standard convex optimization problem. This convex structure ensures computational tractability and allows the problem to be efficiently solved using off-the-shelf convex programming techniques \citep{SCPtutorial}.
\end{enumerate}

Together, these components form a unified and implementable solution for time-sensitive UAV mission planning, with demonstrated feasibility and robustness in simulated urban environments.

\emph{Notation}: We denote by $\mathbb{Z}_{[a,b]}$ the set of all integers in the closed interval $[a,b]$, where $a,b\in \mathbb{R}$ and $a\leq b$.

\section{Problem Formulation}
\subsection{System model}
We consider a UAV operating in a three-dimensional urban environment. The UAV is modeled as a discrete-time system with 3-degree-of-freedom kinematics and control inputs, evolving over a fixed planning horizon of $T$ time steps with a uniform time step $\Delta t$. Let $\mathbf{x}_t = [x_t, y_t, z_t]^\top \in \mathbb{R}^3$ denote the position of the UAV at time step $t$, and let $\mathbf{v}_t = [v^x_t, v^y_t, v^z_t]^\top$ represent its velocity components along each axis.

The discrete-time kinematic model of the UAV is given by:
\begin{equation}\mathbf{x}_{t+1} = \mathbf{x}_t + \mathbf{v}_t \Delta t$$ $$\mathbf{v}_{t+1} = \mathbf{v}_t + \mathbf{a}_t \Delta t\end{equation}
where $\mathbf{a}_t = [a^x_t, a^y_t, a^z_t]^\top \in \mathbb{R}^3$ is the acceleration input applied at time step $t$.

To ensure dynamic feasibility, the UAV is subject to constraints on its speed and acceleration:
\begin{equation}\| [v^x_t, v^y_t]^\top \|_2 \leq v_{\text{max}}, \quad \forall t \in \mathbb{Z}_{[1,T]}$$ $$|a^x_t| \leq a_{\text{max}}, \quad |a^y_t| \leq a_{\text{max}}, \quad |a^z_t| \leq a_{\text{max}}, \quad \forall t\in \mathbb{Z}_{[1,T-1]}\end{equation} 
where $v_{\text{max}}$, $a_{\text{max}}$ are user-defined upper bounds on the UAV’s planar speed and acceleration magnitude, respectively.
The initial position and velocity of the UAV are assumed to be known and fixed:
$$\mathbf{x}_1 = \mathbf{x}_\text{init}, \quad \mathbf{v}_1 = \mathbf{0}$$

\subsection{Task Specification and STL Encoding}
The UAV is tasked with delivering medical supplies—such as blood or plasma units—to a set of geographically distributed hospitals within a fixed planning horizon. Let $\{ \mathbf{h}_1, \mathbf{h}_2, \dots, \mathbf{h}_K \}$ denote the set of hospital locations in 3D space, where each $\mathbf{h}_k \in \mathbb{R}^3$ is a target delivery waypoint. Each hospital $\mathbf{h}_k$ is associated with a temporal delivery window $[\tau_k^\text{start}, \tau_k^\text{end}] \subseteq [1, T]$, during which the UAV must arrive sufficiently close to the destination.

The mission objective is to visit all $K$ hospitals within their respective time windows while respecting dynamic and environmental constraints. Unlike simple waypoint following, this delivery problem imposes temporal validity on each target, making both the spatial and timing aspects critical to mission success. To formally encode this task, we adopt STL as a specification language, which is defined according to:
\begin{equation}
	\varphi:=\pi\ |\ \neg \varphi\ |\ \varphi\land \psi\ | \ \varphi\, \boldsymbol{U}_{[a,b]}\psi,\notag
\end{equation}
where $\pi$ is a predicate associated with a real-valued predicate function $\mu^\pi$; $\varphi$ and $\psi$ are STL formulas; negation and conjunction of formulas are denoted by $\neg$ and $\land$; the until operator is denoted by $\boldsymbol{U}_{[a,b]}$ with $a,b\in [1,T]$. We denote by $(\boldsymbol{x},t)\models \varphi$ a trajectory $\boldsymbol{x}=\{\mathbf{x}_1,\dots,\mathbf{x}_T\}$ satisfies $\varphi$ at time $t$ using the Boolean semantics of STL \citep{Gilpin21smooth}, where
$(\boldsymbol{x},t)\models \pi \Leftrightarrow \mu^\pi(\mathbf{x}_t)\geq 0$. We can evaluate the satisfaction of $\varphi$ quantitatively using a robustness function $\rho^\varphi$ such that $\varphi$ is satisfied if and only if $\rho^\varphi\geq 0$ \citep{Belta19formal}. In particular, we define the eventually operator $\Diamond_{[a, b]} \varphi:=\mathrm{T}\boldsymbol{U}_{[a,b]}\varphi$, where $\mathrm{T}$ is the true predicate. We have $(\boldsymbol{x},t)\models \Diamond_{[a, b]} \varphi$ if $\exists t^\prime\in [t+a,t+b]$ such that $(\boldsymbol{x},t^\prime)\models \varphi$. 
%STL enables the representation of spatial-temporal requirements such as “the UAV must arrive near hospital $\mathbf{h}_k$ between times $\tau_k^\text{start}$ and $\tau_k^\text{end}$.”

\subsubsection{Eventually Operator for Delivery Constraints}

For each hospital $\mathbf{h}_k$, we define the task using the eventually operator $\Diamond_{[\tau_k^\text{start}, \tau_k^\text{end}]} \varphi_k$ , where $\varphi_k$ denotes a spatial predicate indicating proximity to the hospital. Specifically, the predicate $\varphi_k(t)$ is true if the UAV is within a tolerance distance $\varepsilon$ from the delivery point at time $t$:
\begin{equation}\varphi_k(t) := \| \mathbf{x}_t - \mathbf{h}_k \|_2 \leq \varepsilon,\ t\in\mathbb{Z}_{[\tau_k^\text{start},\tau_k^\text{end}]}\end{equation}
The temporal logic specification requires that for each hospital, the predicate must be satisfied at least once within its time window:
\begin{equation}(\boldsymbol{x},0)\models\Diamond_{[\tau_k^\text{start}, \tau_k^\text{end}]} \varphi_k, \quad \forall k \in \mathbb{Z}_{[1,K]}\end{equation}

\subsubsection{Robustness Semantics}

To make STL compatible with continuous optimization, we adopt its quantitative semantics (robustness measure). The robustness value $\rho_k(t)$ for predicate $\varphi_k(t)$ is defined as:
\begin{equation}\rho_k(t) = \varepsilon - \| \mathbf{x}_t - \mathbf{h}_k \|_2,\ t\in\mathbb{Z}_{[\tau_k^\text{start},\tau_k^\text{end}]}\end{equation}

Higher values of $\rho_k(t)$ indicate stronger satisfaction (i.e., the UAV is well within the target region), while negative values indicate violation. The robustness of the full temporal operator is then defined as the maximum robustness over the time window:
\begin{equation}\rho_k^{\Diamond} = \max\{ \rho_k(t) , t\in\mathbb{Z}_{[\tau_k^\text{start},\tau_k^\text{end}]}\},\ \forall k \in \mathbb{Z}_{[1,K]}.
\end{equation} 
Enforcing STL satisfaction therefore amounts to requiring $\rho_k^{\Diamond} \geq 0$ for all hospitals.

\subsection{Obstacle Representation}
The operating environment is assumed to be an urban area populated with static obstacles such as buildings. To facilitate collision checking, each obstacle is modeled as a three-dimensional, axis-aligned rectangular prism, commonly referred to as an Axis-Aligned Bounding Box (AABB). Assume $M$ obstacles exist in the environment, let $\mathcal{O}_j \subset \mathbb{R}^3$ denote the region occupied by the $j$-th obstacle. This region is defined as:
\begin{equation}\mathcal{O}_m \!=\! \left\{ \mathbf{p}\! =\! \begin{bmatrix} p_x \\ p_y \\ p_z \end{bmatrix} \!\in\! \mathbb{R}^3\! \;\middle|\; \begin{aligned} & x_m^{\min} \leq p_x \leq x_m^{\max}, \\ & y_m^{\min} \leq p_y \leq y_m^{\max}, \\ & z_m^{\min} \leq p_z \leq z_m^{\max} \end{aligned} \right\},\forall m\!\in\!\mathbb{Z}_{[1,M]}\end{equation} 
where $x_m^{\min}, x_m^{\max}, y_m^{\min}, y_m^{\max}, z_m^{\min}, z_m^{\max}$ define the lower and upper bounds of the obstacle along each axis. This representation enables efficient geometric reasoning and allows for straightforward computation of signed distances between the UAV and obstacle boundaries.

To ensure collision-free operation, the UAV’s position $\mathbf{x}_t$ at each time step must remain outside all defined obstacle regions:
\begin{equation}\mathbf{x}_t \notin \mathcal{O}_m, \ \forall m\in\mathbb{Z}_{[1,M]}, \ \forall t \in \mathbb{Z}_{[1,T]}\end{equation} 

\subsection{Objective Function}
The objective function is designed to guide the UAV toward fulfilling the delivery mission while ensuring smooth and safe motion. It consists of three main components: (i) maximization of STL robustness, (ii) penalization of control effort, and (iii) penalization of proximity to obstacles. Each term is weighted to balance task completion against energy efficiency and safety.

To promote task success and encourage early satisfaction of delivery objectives, we include a term that maximizes the robustness of the temporal logic specification:
\begin{equation}J_{\text{STL}} = - \sum_{k=1}^{K} \rho_k^{\Diamond}\end{equation} 
The negative sign allows this term to be minimized in the convex optimization framework, effectively maximizing the STL robustness values.

To ensure that the resulting trajectory is dynamically smooth and energy-efficient, we penalize excessive acceleration. The control effort term is defined as the squared $\ell_2$-norm of the acceleration inputs:
\begin{equation}J_{\text{control}} = \sum_{t=1}^{T-1} \| \mathbf{a}_t \|_2^2\end{equation} 
This term discourages sharp or abrupt changes in the UAV's velocity profile.

Although obstacle avoidance is enforced through constraints, we include an obstacle proximity penalty term that discourages the UAV from flying too close to obstacle boundaries. Following \citep{Liu18convex}, a signed distance function $\phi_m(\mathbf{x}_t)$ is defined for each obstacle $\mathcal{O}_m$, and the proximity penalty is computed using a hinge loss:
\begin{equation}J_{\text{obs}} = \sum_{m=1}^{M} \sum_{t=1}^{T} \max(0, d_{\text{safe}} - \phi_m(\mathbf{x}_t))\end{equation} 
where $d_{\text{safe}}$ is a user-defined safety margin. This term encourages conservative, safe separation from obstacles.

The total cost function is expressed as a weighted combination of the three components:
\begin{equation}J = w_1 J_{\text{STL}} + w_2 J_{\text{control}} + w_3 J_{\text{obs}}\end{equation} 
where $w_1, w_2, w_3 > 0$ are scalar weights used to balance mission satisfaction, control smoothness, and safety. These weights are tuned empirically to ensure reliable delivery while avoiding overly aggressive or conservative behavior.

\section{Optimization Framework}
This section presents the reformulation of the trajectory optimization problem into a tractable convex framework. A key challenge in encoding STL constraints lies in the non-smooth and non-convex nature of temporal operators, particularly the max operation in robustness evaluation. Directly optimizing such specifications is computationally expensive and often unsuitable for real-time planning.

To address this, we approximate the STL robustness semantics using smooth, recursive sub-dynamics, enabling the formulation of differentiable constraints compatible with convex programming. In parallel, obstacle avoidance typically non-convex is handled through a linearized approximation around a reference trajectory. This results in a sequence of convex subproblems constructed within a convex feasible set, ensuring both safety and tractability. Together, these approximations allow the entire mission planning problem to be expressed as a standard convex optimization program that remains efficient to solve while faithfully respecting high-level temporal and spatial requirements.

\subsection{Convex Reformulation of STL Constraints}
We introduce a smooth approximation (or sub-dynamic recursion) to avoid non-smooth and non-convex max operators in optimization. A set of auxiliary variables $\mu_k(t)$, $t\in\mathbb{Z}_{[\tau_k^\text{start},\tau_k^\text{end}]}$ that recursively track the maximum robustness value are introduced:
\begin{equation}
	\begin{aligned}
		\mu_k(t) &= \max(\mu_k(t-1), \rho_k(t)) \\
		&\approx G(\mu_k(t-1),\rho_k(t),\alpha)\\
		&:=\frac{1}{2} \left( \mu_k(t-1)\! +\! \rho_k(t)\! +\! \sqrt{(\mu_k(t-1)\! -\! \rho_k(t))^2\! +\! \alpha^2} \right)
	\end{aligned}\notag
\end{equation}
where $\alpha > 0$ is a smoothing parameter. This approximation is then linearized about reference trajectories $\bar{\mu}_k$ and $\bar{\rho}_k$, yielding:
\begin{equation}
	\begin{aligned}
		&\mu_k(t)=G(\tilde \mu_k(t-1),\tilde\rho_k(t),\alpha)\\
		&+\frac{1}{2}\!\left(\!1\!+\!\frac{\tilde \mu_k(t-1)-\tilde\rho_k(t)}{\sqrt{\!(\tilde \mu_k(t\!-\!1)\! -\! \tilde\rho_k(t))^2\! +\! \alpha^2}}\!\right)(\mu_k(t\!-\!1)-\tilde \mu_k(t\!-\!1))\\
		&+\frac{1}{2}\!\left(\!1\!+\!\frac{\tilde\rho_k(t)-\tilde \mu_k(t-1)}{\sqrt{\!(\tilde \mu_k(t\!-\!1)\! -\! \tilde\rho_k(t))^2\! +\! \alpha^2}}\!\right)(\rho_k(t)-\tilde \rho_k(t)).
	\end{aligned}\notag
\end{equation}
The final STL satisfaction condition can be compactly expressed as:
$$\mu_k(\tau_k^\text{end}) \geq 0, \quad \forall  k \in \mathbb{Z}_{[1,K]}$$ 
These constraints are integrated into the overall trajectory optimization problem, allowing the UAV to satisfy delivery objectives in a differentiable and optimization-friendly manner.
For each hospital $k$, the auxiliary robustness variable $\mu_k(t)$ tracks the satisfaction level of the STL eventually constraint over time. 
We aim to maximize the robustness, which corresponds to the strongest satisfaction of the time-windowed goal
$$J_{\text{STL}}^c = - \sum_{k=1}^{K} \mu_k(\tau_k^{\text{end}})$$

Reformulating the original STL constraints into a smooth recursive structure not only avoids the computational burden of non-convex optimization but also ensures compatibility with convex solvers. This allows STL satisfaction to be enforced using standard tools without requiring integer variables or case-based logic. More importantly, the recursive formulation preserves the temporal semantics of the delivery task—ensuring that the UAV consistently increases its satisfaction measure within the specified time window—while enabling gradient-based optimization methods to efficiently find feasible and robust trajectories. This smooth convex approximation serves as the foundation for integrating STL specifications into practical trajectory planning.

\subsection{CFS for Obstacle Avoidance}

In order to address the non-convexity of obstacle avoidance constraints, we adopt the Convex Feasible Set methodology, as originally proposed in \citep{Liu18convex}. The key idea is to transform the non-convex constraint set into a sequence of convex approximations centered around a reference trajectory. This allows the original trajectory optimization to be solved via an iterative convex programming procedure.

Let each obstacle $\mathcal{O}_m \subset \mathbb{R}^3$ be represented implicitly by a signed distance function $\phi_m(\mathbf{x})$, such that:
$$\phi_m(\mathbf{x}) \geq 0 \iff \mathbf{x} \notin \mathcal{O}_m$$

Assuming $\phi_m$ is semi-convex and piecewise smooth (as supported by signed or directional distance functions), the feasible region for the UAV is:
$$\Gamma := \bigcap_{m=1}^M \left\{ \mathbf{x} \in \mathbb{R}^3 : \phi_m(\mathbf{x}) \geq 0 \right\}$$
At each iteration $i$, given a reference trajectory $\bar {\mathbf{x}}_t:=\mathbf{x}^{(i-1)}_t$, we construct a linearized convex constraint via the first-order Taylor expansion of $\phi_m$:
\begin{equation}
	\begin{aligned}
		\phi_m^c(\mathbf{x}_t):=\phi_m(\bar {\mathbf{x}}_t)+\nabla  \phi_m(\bar {\mathbf{x}}_t)(\mathbf{x}_t-\bar{\mathbf{x}}_t )\geq 0, \\\forall m\in\mathbb{Z}_{[1,M]},\ \forall t\in \mathbb{Z}_{[1,T]},\label{convex_poly}
	\end{aligned}\notag
\end{equation}
where $\nabla$ denotes the subgradient that minimizes the cost function in the feasible set. This defines a local convex feasible set $\mathcal{F}^{(i)} \subseteq \Gamma$, which is used to solve the next convex subproblem. Iteration proceeds until convergence in cost or trajectory. Analogously, the proximity penalty is convexified to:
$$J_{\text{obs}}^c = \sum_{m=1}^{M} \sum_{t=1}^{T} \max(0, d_{\text{safe}} - \phi_m^c(\mathbf{x}_t))$$

Compared to conventional non-convex formulations, the CFS approach offers several advantages. First, it guarantees feasibility at each iteration as long as the reference trajectory is feasible. Second, it exploits the geometric structure of the motion planning problem, allowing for fast convergence with minimal per-iteration computation. Finally, the formulation is compatible with standard convex solvers, making it well-suited for integration with the STL-based delivery framework developed in this paper.

\subsection{The Convex Optimization Problem}
Through the series of reformulations introduced in previous sections, the original trajectory planning problem—characterized by non-smooth temporal logic constraints and non-convex obstacle regions—is transformed into a convex optimization problem that is both tractable and solver-friendly. Specifically, the non-smooth STL constraints are approximated using recursively defined smooth sub-dynamics, allowing STL robustness to be embedded as convex inequality constraints. Meanwhile, obstacle avoidance is handled using linearized constraints around a reference trajectory, derived from signed distance functions, ensuring convex feasibility at each iteration.

The complete convex optimization problem is summarized as follows:
\begin{equation}
	\begin{aligned}
		\min\limits_{} &\quad w_1 J_{\text{STL}}^c + w_2 J_{\text{control}} + w_3 J_{\text{obs}}^c  \\
		\text {s.t.} 
		&\quad \mathbf{x}_{t+1} = \mathbf{x}_t + \mathbf{v}_t \Delta t\\
		&\quad \mathbf{v}_{t+1} = \mathbf{v}_t + \mathbf{a}_t \Delta t
	\end{aligned}\notag
\end{equation}
\begin{equation}
	\begin{aligned}
 &\| [v^x_t, v^y_t]^\top \|_2 \leq v_{\text{max}}, \ \forall t \in \mathbb{Z}_{[1,T]}\\
		&\quad |a^x_t| \leq a_{\text{max}}, \ |a^y_t| \leq a_{\text{max}}, \ |a^z_t| \leq a_{\text{max}}, \ \forall t\in \mathbb{Z}_{[1,T-1]}\\
		&\quad \mathbf{x}_1 = \mathbf{x}_\text{init}, \quad \mathbf{v}_1 = \mathbf{0} \\
		&\quad \mu_k(\tau_k^\text{end}) \geq 0, \ \forall  k \in \mathbb{Z}_{[1,K]}\\
		&\quad \phi_m^c(\mathbf{x}_t)\geq 0, \ \forall m\in\mathbb{Z}_{[1,M]},\ \forall t\in \mathbb{Z}_{[1,T]}
	\end{aligned}\notag
\end{equation}
This formula consists of convex dynamics, convexified task satisfaction constraints, and locally convex approximations of non-convex environment constraints. All terms in the objective function and all constraints are either affine or convex, making the overall formulation compatible with standard convex solvers. This enables efficient and reliable computation of safe, dynamically feasible UAV trajectories that respect mission timing and spatial requirements. The proposed convex optimization algorithm for UAV trajectory planning under STL and collision avoidance constraints is summarized in Algorithm \ref{Alg1}.

\begin{algorithm}[htb]
	\caption{The proposed algorithm}\label{Alg1}
	\begin{algorithmic}[1]
		\Statex \textbf{Input:} Waypoints $\{ \mathbf{h}_1, \mathbf{h}_2, \dots, \mathbf{h}_K \}$ and the associated time windows $[\tau_k^\text{start}, \tau_k^\text{end}]$ for all $k\in \mathbb{Z}_{[1,K]}$. Positive weights $w_1$, $w_2$, and $w_3$. Initial reference trajectory $\bar {\mathbf{x}}$.
		\State Compute initial reference $\bar{\mu}_k$ and $\bar{\rho}_k$ according to 
            \begin{equation}
			\bar{\mu}_k(t)=\bar\rho_k(t) = \varepsilon - \| \bar{\mathbf{x}}_t - \mathbf{h}_k \|_2,\ t\in\mathbb{Z}_{[\tau_k^\text{start},\tau_k^\text{end}]},\ k\in\mathbb{Z}_{[1,K]}\notag
		\end{equation}
		\While {not converged}
		\State Formulate the convex optimization problem following the procedures in Sec. III.
		\State Solve the problem to obtain a new solution $\{\mathbf{x},\mathbf{v},\mathbf{a},\boldsymbol{\rho},\boldsymbol{\mu}\}$, where $\boldsymbol{\rho}$ and $\boldsymbol{\mu}$ collects all the $\rho_k$ and $\mu_k$ auxiliary variables.
		\State Update reference trajectory by $\bar{\mathbf{x}}=\mathbf{x}$, $\bar{\boldsymbol{\rho}}=\boldsymbol{\rho}$, and $\bar{\boldsymbol{\mu}}=\boldsymbol{\mu}$.
		\EndWhile
		\Statex \textbf{Output:} The desired trajectory $\{\mathbf{x},\mathbf{v},\mathbf{a},\boldsymbol{\rho},\boldsymbol{\mu}\}$.
	\end{algorithmic}
\end{algorithm}

\section{Simulation Result}
\subsection{Parameter Specification}
The proposed trajectory optimization framework is tested in a simulated three-dimensional urban environment. All simulations are implemented in MATLAB using the convex optimization toolbox (CVX) \citep{cvx} over a discretized time horizon of $T = 30$ time steps with a uniform interval $\Delta t = 1.0$ seconds.

The UAV begins at the fixed initial position $\mathbf{x}_1 = [0,\; 0,\; 5]^\top$, which
represents a rooftop launch pad at 5 meters above ground. The UAV's initial velocity is zero in all directions. Its motion follows the discrete-time kinematic model defined in Section 3.1, with maximum planar speed $v_{\text{max}} = 5 \,\text{m/s}$, and acceleration bounds $a_{\text{max}} = 2 \,\text{m/s}^2$ in all directions.

The UAV is required to deliver medical supplies to three hospitals, each defined by a 3D location and a specific delivery time window (shown in Table I):
\begin{table}[htbp]
	\centering
	\caption{Delivery Targets}
	\label{tab:parameters}
	\begin{tabular}{lcc}
		\hline
		\hline
		\textbf{Hospital} & \textbf{Coordinates} & \textbf{Time Window} \\ 
		&$(x, y, z)$&$[\tau^{\text{start}}, \tau^{\text{end}}]$\\\hline
		$\mathbf{h}_1$ & $[17,\; 18,\; 5]$ & $[4, 15]$  \\
		$\mathbf{h}_2$ & $[42,\; 28,\; 5]$ & $[10, 22]$ \\
		$\mathbf{h}_3$ & $[56,\; 20,\; 5]$ & $[16, 30]$ \\
		\hline
		\hline
	\end{tabular}
\end{table}

Each delivery is considered successful if the UAV enters an $\varepsilon = 0.2 \,\text{m}$ radius of the hospital location during the corresponding time window, as defined by the STL eventually constraint.

The simulation environment includes three static cuboid obstacles, representing buildings placed to challenge the trajectory planning. Each obstacle is defined as an axis-aligned 3D box using its lower and upper corner coordinates:
\begin{table}[htbp]
	\centering
	\caption{Obstacle Specification}
	\label{tab:parameters}
	\begin{tabular}{lccc}
		\hline
		\hline
		\textbf{Obstacle} & $x$-\textbf{Range}  & $y$-\textbf{Range} & $z$-\textbf{Range}\\ \hline
		$\mathcal{O}_1$ & $[10, 15]$ & $[5, 15]$  & $[0, 15]$\\
		$\mathcal{O}_2$ & $[30, 40]$ & $[20, 30]$ & $[0, 20]$ \\
		$\mathcal{O}_3$ & $[45, 55]$ & $[5, 15]$ & $[0, 20]$ \\
		\hline
		\hline
	\end{tabular}
\end{table}

The optimization is solved over $T = 30$ discrete steps using CVX. An iterative convexification procedure is used, where a reference trajectory is refined through multiple rounds of convex optimization with obstacle-aware linearization. 

\subsection{Results and Discussions}
To validate the proposed convex optimization framework, we conduct simulations in the urban environment described previously. The results are summarized in Fig. 1 through Fig. 4, demonstrating the effectiveness, convergence behavior, and task satisfaction capability of the method.

\subsubsection{Trajectory Quality and Task Completion}

Fig. 1 compares the optimized UAV trajectory obtained using the proposed convex approach (orange) against that generated by a standard Sequential Quadratic Programming (SQP) solver (blue) \citep{Gill05snopt}. Both 3D and top-down views are shown. The convex method successfully produces a smooth, dynamically feasible trajectory that visits all three hospitals while respecting the obstacle constraints and STL-defined time windows. In contrast, the SQP-based trajectory visibly deviates near the second and third hospitals and slightly overshoots the time window boundaries, indicating a lack of robustness in satisfying temporal logic constraints.
\begin{figure}[!htb]
	\centering
	\includegraphics[width=0.7\linewidth]{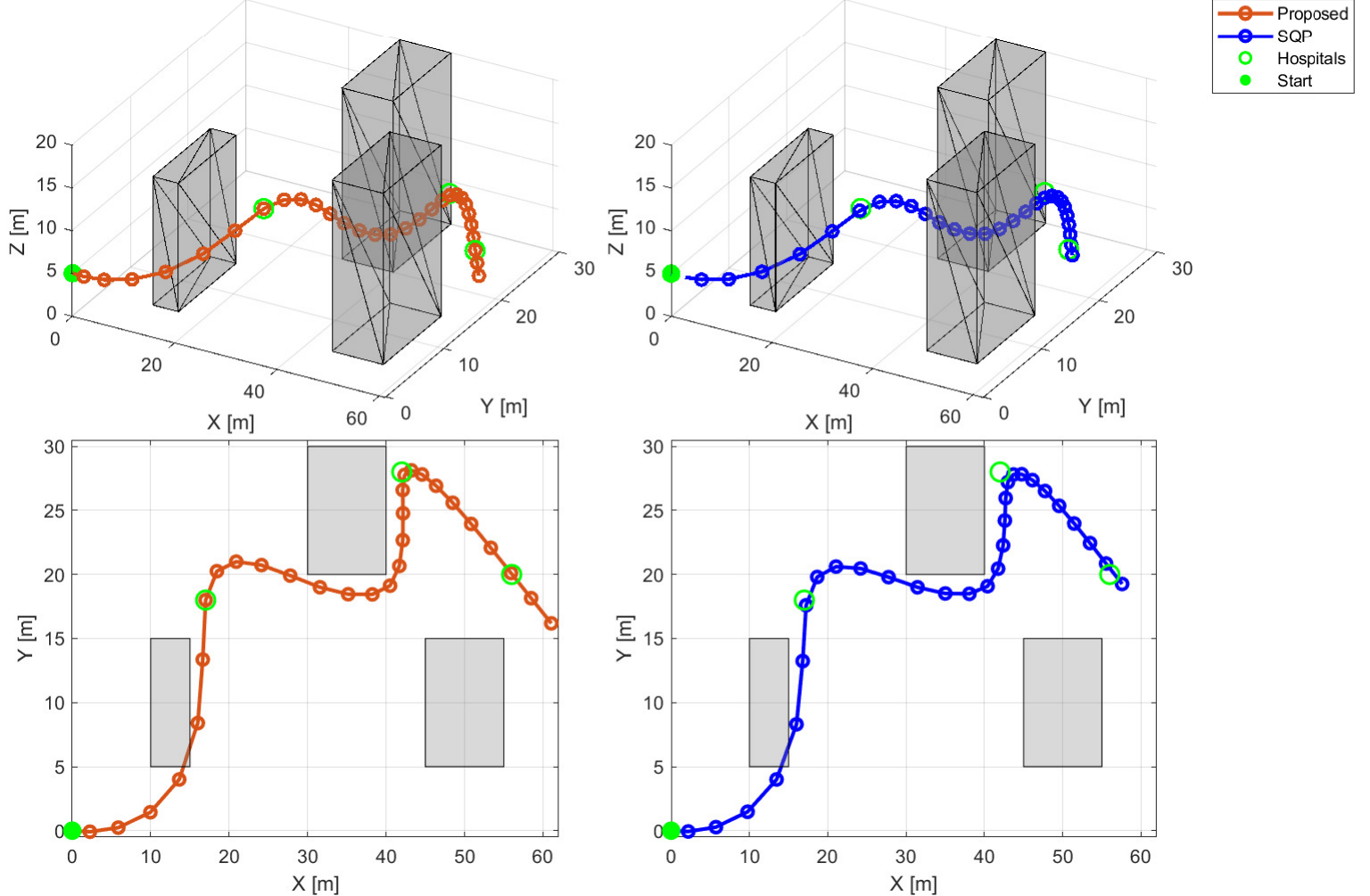}
	\caption{Trajectory results}
	\label{fig:enter-label}
\end{figure}

The proposed approach maintains appropriate clearance from obstacles, resulting in a trajectory that is both safe and executable. The path transitions are continuous and smooth, reflecting the implicit control effort regularization in the objective function.

\subsubsection{Fast Convergence and Optimization Efficiency}
The convergence of the optimization algorithm is illustrated in Fig. 2, which plots the total objective value as a function of the outer-loop iteration number. The objective drops rapidly in the first few iterations and stabilizes after approximately 5 rounds. This behavior indicates that the proposed STL sub-dynamics and CFS update mechanism are effective at quickly steering the solution toward an optimal region. Most of the performance gain is achieved in the first few iterations, enabling the method to support early termination if time-critical constraints are imposed.
\begin{figure}[!htbp]
	\centering
	\includegraphics[width=0.65\linewidth]{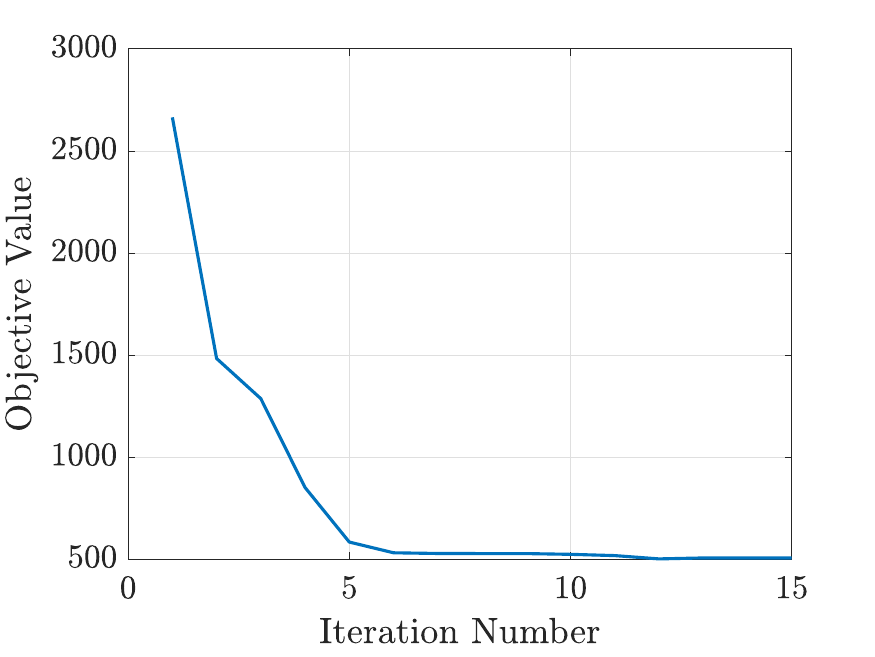}
	\caption{Objective function value history}
	\label{fig:enter-label}
\end{figure}
\begin{figure}[!htbp]
	\centering
	\includegraphics[width=0.7\linewidth]{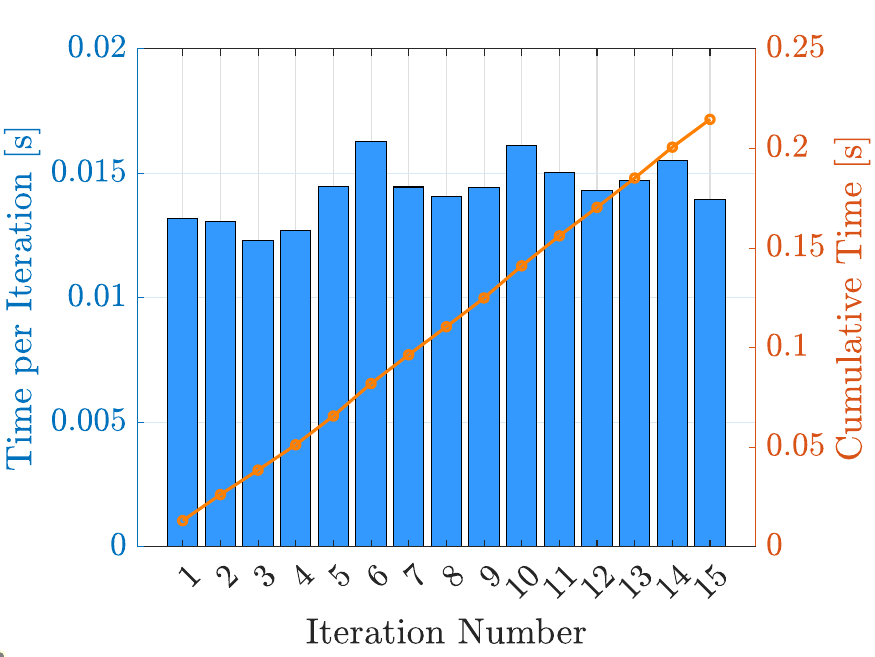}
	\caption{Computational performance}
	\label{fig:enter-label}
\end{figure}

Fig. 3 further confirms the computational efficiency of the approach. The average CPU time per iteration is consistently below 0.02 seconds, with all iterations completing in under 0.22 seconds cumulatively. This lightweight computational profile makes the proposed method well-suited for real-time or onboard applications where computation resources are limited.

\subsubsection{Temporal Logic Satisfaction and Robustness}
To assess how well the trajectory satisfies the STL temporal constraints, Fig. 4 shows the evolution of the robustness values 
$\mu_{k}$(t) for each of the three hospital goals over the planning horizon. Each curve corresponds to the quantitative satisfaction level of the eventually condition associated with a hospital target.
\begin{figure}[!htbp]
	\centering
	\includegraphics[width=0.6\linewidth]{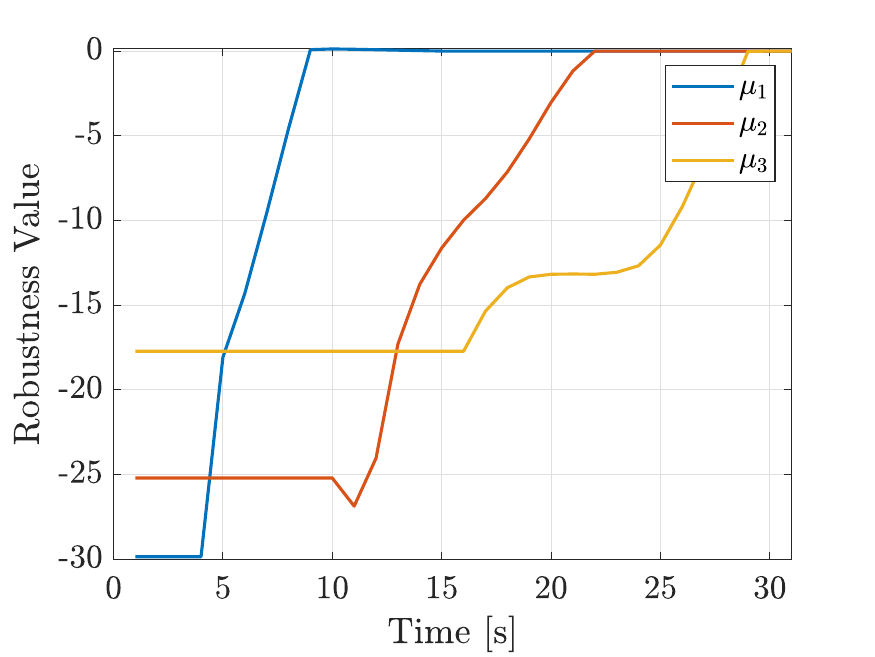}
	\caption{STL function history}
	\label{STL function history}
\end{figure}

The robustness functions exhibit a sharp rise as the UAV approaches each hospital within its time window, eventually reaching zero and staying flat afterward. This behavior confirms that the UAV not only enters the goal region but does so within the prescribed temporal interval and with a sufficient safety margin. The positive slope of each $\mu_{k}$(t) curve further suggests that the planner actively improves satisfaction over time rather than passively meeting minimal requirements.

\subsubsection{Comparative Discussion and Practical Implications}
Compared to the SQP baseline, the proposed method yields more consistent satisfaction of STL goals, better obstacle margin handling, and lower trajectory curvature. Additionally, the convex formulation avoids local minima issues that often arise in nonlinear optimization, as evidenced by the smoother and more accurate trajectory shapes in Fig. 1.

The iterative structure of the method, combined with convex programming at each stage, enables robust and predictable convergence even from a simple straight-line initialization. These properties make the approach particularly suitable for deployment in safety-critical scenarios such as UAV-based medical delivery, where mission success and timing compliance are essential.

%\subsection{Applications in Smart Systems and Autonomous Networks}

\section{Conclusion}
This paper presents a convex trajectory optimization framework for UAV-based medical delivery missions with spatial and temporal constraints. By reformulating STL specifications into smooth sub-dynamics and applying convex approximations for obstacle avoidance, the method transforms a non-convex planning problem into a standard convex program solvable by efficient optimization tools. Simulation results show that the proposed approach produces dynamically feasible, obstacle-free trajectories that successfully satisfy all delivery deadlines, outperforming conventional nonlinear methods in both robustness and trajectory quality. The algorithm also exhibits fast convergence and low computational overhead, making it well-suited for real-time implementation. Future work will extend the approach to multi-agent coordination and dynamic, uncertain environments.

\printcredits

%% Loading bibliography style file
% \bibliographystyle{model1-num-names}
\bibliographystyle{cas-model2-names}

% Loading bibliography database
\bibliography{Review}

%\vskip3pt

%\bio{}
%Author biography without author photo.
%Author biography. Author biography. Author biography.
%Author biography. Author biography. Author biography.
%Author biography. Author biography. Author biography.
%Author biography. Author biography. Author biography.
%Author biography. Author biography. Author biography.
%Author biography. Author biography. Author biography.
%Author biography. Author biography. Author biography.
%Author biography. Author biography. Author biography.
%Author biography. Author biography. Author biography.
%\endbio

%\bio{figs/pic1}
%Author biography with author photo.
%Author biography. Author biography. Author biography.
%Author biography. Author biography. Author biography.
%Author biography. Author biography. Author biography.
%Author biography. Author biography. Author biography.
%Author biography. Author biography. Author biography.
%Author biography. Author biography. Author biography.
%Author biography. Author biography. Author biography.
%Author biography. Author biography. Author biography.
%Author biography. Author biography. Author biography.
%\endbio
%
%\bio{figs/pic1}
%Author biography with author photo.
%Author biography. Author biography. Author biography.
%Author biography. Author biography. Author biography.
%Author biography. Author biography. Author biography.
%Author biography. Author biography. Author biography.
%\endbio

%\end{sloppypar}
\end{document}